\documentclass[12pt]{article}
\usepackage{epsfig}

\def\beq{\begin{equation}}
\def\eeq#1{\label{#1}\end{equation}}
\def\eeqn{\end{equation}}

\def\beqa{\begin{eqnarray}}
\def\eeqa#1{\label{#1}\end{eqnarray}}
\def\eeqan{\end{eqnarray}}

\let\bar=\overbar

\def\Dslash{\not{\hbox{\kern-4pt $D$}}}
\def\dslash{\not{\hbox{\kern-2pt $\del$}}}

\def\msb{{\bar{\ssstyle M \kern -1pt S}}}

\def\Title#1{\begin{center} {\Large {\bf #1} } \end{center}}

\begin{document}

\Title{The observational appearance of strange stars}

\bigskip\bigskip


\begin{raggedright}  

{\it Vladimir V. Usov\index{Usov, V.V.}\\
Department of Condensed Matter Physics\\
Weizmann Institute of Science\\
Rehovot 76100, ISRAEL}
\bigskip\bigskip
\end{raggedright}

\section{Introduction}
Strange stars that are entirely made of strange quark matter (SQM)
have been long ago proposed as an alternative to neutron stars
(e.g., \cite{W84,AFO86}). The possible existence of strange 
stars is a direct consequence of the conjecture that
SQM composed 
of roughly equal numbers of up, down, and strange quarks 
plus a smaller numbers of electrons (to neutralize the electric
charge of the quarks) may be the 
absolute ground state of the strong interaction, i.e., absolutely
stable with respect to $^{56}$Fe \cite{W84,B71}. 
The bulk properties (size, moment of inertia, etc.) of models 
of strange and neutron stars in the observed mass range $(1< M/M_\odot
<2)$ are rather similar, and it is very difficult to discriminate 
between strange and neutron stars \cite{G97}. 

SQM with the density 
of $\sim 5\times 10^{14}$ g~cm$^{-3}$ might exist
up to the surface of strange stars \cite{AFO86,G97}. 
Such a bare strange star differs qualitatively from a neutron star 
which has the density at the stellar surface (more exactly at the
stellar photosphere) of about $0.1-1~{\rm g~cm}^{-3}$. This opens 
observational possibilities to distinguish bare strange stars from 
neutron stars.

\section{Thermal emission from bare strange stars}

At the bare SQM surface of a strange star the density changes 
abruptly from $\sim 5\times 10^{14}$ g~cm$^{-3}$ to zero. The thickness
of the SQM surface is about $1~{\rm fm}=10^{-13}$~cm, which is a 
typical strong interaction length scale. Since SQM at the surface of 
a bare strange star is bound via strong
interaction rather than gravity, such a star is not subject to
the Eddington limit in contrast to a neutron star \cite{AFO86,U01a}. 
Below, we discuss the thermal emission of photons and $e^+e^-$ 
pairs from the SQM surface of a hot bare strange star.

\subsection{Emission of photons}

Hot SQM is filled with electromagnetic waves  
in thermodynamic equilibrium with quarks.  The dispersion 
relation of these waves may be written in the following simple form 
$\omega^2=\omega_{\rm p}^2+k^2c^2$, where $\omega$ is the frequency 
of electromagnetic waves, $k$ is their wavenumber, and $\omega_{\rm p}$
is the plasma frequency of quarks \cite{AFO86}. This equation is 
the familiar dispersion relation for a plasma, 
and its conventional interpretation may be applied to SQM. Propagating 
modes exist only for $\omega > \omega_{\rm p}$. Therefore,
there is the lower limit on the energy of 
electromagnetic photons that are in thermodynamic equilibrium with 
quarks, $\varepsilon_\gamma =\hbar \omega > \hbar \omega_{\rm p}\simeq 
18.5(n_{\rm b}/n_0)^{1/3}$ MeV, where $n_{\rm b}$ is the baryon number 
density of SQM, and $n_0=0.17$~fm$^{-3}$ is normal 
nuclear matter density. At the
SQM surface where the pressure is zero, we expect
$n_{\rm b}\simeq (1.5-2)n_0$ and $\hbar \omega_{\rm p}\simeq 20-25$ MeV.
i.e., the spectrum of thermal equilibrium 
photons radiated from the bare SQM surfaces of strange stars 
is very hard, $\varepsilon_\gamma > \hbar \omega_{\rm p}\simeq 
20-25$ MeV \cite{AFO86}.

The energy flux emitted from the unit surface of SQM in thermal 
equilibrium photons is \cite{U01a,CHS91}

\begin{equation}
F_{\rm eq}={\hbar \over c^2}\int_{\omega_{\rm p}}^{\infty}d\omega\,
{\omega\, (\omega^2 -\omega_{\rm p}^2)\,g(\omega )\over
{\exp\, (\hbar \omega/k_{_{\rm B}}T_{_{\rm S}})-1}}\,,
\end{equation}

\noindent
where 

\begin{equation}
g(\omega )={1\over 2\pi^2}\int_{0}^{\pi/2}d\vartheta\,
\sin\vartheta\,\cos \vartheta\, D(\omega, \vartheta)\,,
\end{equation}

\noindent
$k_{_{\rm B}}$ is the Boltzmann constant, $T_{_{\rm S}}$ is 
the surface temperature,
$D(\omega,\vartheta)$ is the coefficient of radiation transmission
from SQM to vacuum, $D=1-(R_\perp +R_\parallel)/2$, 
and

\begin{equation}
R_\perp ={\sin^2(\vartheta -\vartheta_0)\over \sin^2(\vartheta 
+ \vartheta_0)}\,,\,\,\,\,\,
R_\parallel={\tan^2(\vartheta -\vartheta_0)\over \tan^2(\vartheta 
+ \vartheta_0)}\,,\,\,\,\,\,
\vartheta_0= \arcsin\, \left[\,\sin \vartheta\,\sqrt{1-
\left({\omega_{\rm p}\over 
\omega}\right)^2}\,\right].
\end{equation}

\begin{figure}[htb]
\begin{center}
\epsfig{file=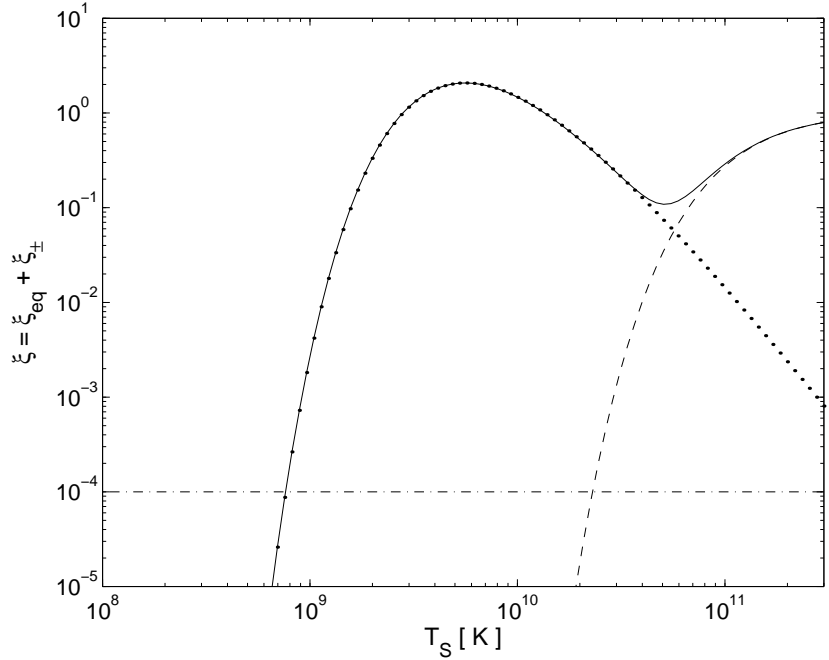,height=3.5in}
\caption{Total emissivity of bare SQM surface
(solid line), which is the sum of emissivities 
in equilibrium photons (dashed line) and $e^+e^-$ pairs 
(dotted line), divided by the blackbody emissivity,
$\xi=\xi_{\rm eq}+\xi_\pm$.
The value of $\xi$ is valid at least for the surface temperature
$T_{_{\rm S}}\geq 8\times 10^8$~K when $\xi$ is more than
the upper limit on $\xi_{\rm neq}$, $\xi_{\rm neq}\leq 10^{-4}$,
which is shown by the dot-dashed line.}
\label{fig:ksi}
\end{center}
\end{figure}

\noindent 
Figure~1 shows the ratio of the equilibrium photon emissivity
of the bare SQM surface to the blackbody
surface emissivity, $\xi_{\rm eq} = F_{\rm eq}/F_{_{\rm BB}}$, where 
$F_{_{\rm BB}}=\sigma T_{_{\rm S}}^4$, and $\sigma$ is the
Stefan-Boltzmann constant. From Figure 1 we can see that
at $T{_{\rm S}}\ll\hbar \omega_{\rm p}/k_{_{\rm B}}\sim 10^{11}$~K
the equilibrium photon radiation from the bare surface of 
a strange star is very small, compared to the blackbody one.

Low energy photons ($\varepsilon_\gamma < \hbar \omega_{\rm p}$)
may leave SQM if they are produced by a
non-equilibrium process in the surface layer with the thickness 
of $\sim c/\omega_{\rm p}\simeq 10^{-12}$~cm. The upper limit on 
the emissivity of SQM in non-equilibrium photons at low energies is 
$\xi _{\rm neq}=F_{\rm neq}/F_{_{\rm BB}}\leq 10^{-4}$ \cite{CHS91}. 

\subsection{Emission of $e^+e^-$ pairs}

It was pointed out \cite{U98} that the bare surface of a hot
strange star may be a powerful source of $e^+e^-$ pairs which are
created in an extremely strong electric field at the quark surface and
flow away from the star. The electric field is generated because there 
are electrons with the density $n_{\rm e}
\simeq (10^{-3}-10^{-4})n_{\rm b}$ in
SQM to neutralize the electric charge of the quarks (e.g., 
\cite{AFO86,G97}). The point is that the electrons, being bound to 
SQM by the electromagnetic interaction alone, are able
to move freely across the SQM surface, but clearly cannot move to
infinity because of the bulk electrostatic attraction to the quarks.
The electron distribution extends up to $\sim 10^3$ fm above the quark
surface, and a strong electric field is generated in the  
surface layer to prevent the electrons from escaping to infinity, 
counterbalancing the degeneracy  and thermal pressure. The typical 
magnitude of the electric field at the SQM surface is
$\sim 5\times 10^{17}$ V~cm$^{-1}$ \cite{AFO86}. 
This field is a few ten times higher than the critical field  
$E_{\rm cr}=m^2c^3/e\hbar \simeq 1.3\times 10^{16}\,\,{\rm V~cm}
^{-1}$ at which vacuum is unstable to creation of $e^+e^-$ pairs.
In such a strong electric field, $E\gg E_{\rm cr}$, in vacuum, 
the pair creation rate is extremely high,
$W_\pm \simeq 1.7\times 10^{50}({E/ E_{\rm cr}})^2\,\,{\rm cm}^{-3}\,
{\rm s}^{-1}$. At $E\simeq 5\times 10^{17}$ V~cm$^{-1}$, we have 
$W_\pm\simeq 2.5\times 10^{53}$ 
cm$^{-3}$~s$^{-1}$. The high-electric-field region is, however, 
not a vacuum. The electrons present fill up states into which
would-be-created electrons have to go. This reduces the
pair-creation rate from the vacuum value. At zero temperature 
the process of pair creation is suppressed altogether because 
there is no free levels for electrons to be created \cite{U98}.

At finite temperatures, $T_{_{\rm S}}>0$, in thermodynamical 
equilibrium electronic states are only partly filled, and pair 
creation by the Coulomb barrier 
becomes possible. Since the rate of pair production when electrons 
are created into the empty states is extremely high, the empty
states below the pair creation threshold, $\varepsilon\leq 
\varepsilon_{_{\rm F}}-2m_{\rm e}c^2$, are occupied by created 
electrons almost instantly, where 
$\varepsilon_{_{\rm F}}=\hbar c(\pi^2n_{\rm e})^{1/3}\simeq 20$~MeV 
is the Fermi 
energy of electrons in SQM, and $m_{\rm e}$ is the electron mass
\cite{U98}. Then, the rate of pair creation by
the  Coulomb barrier is determined by the process of thermalization
of electrons which favors the empty-state production below the
pair creation threshold. The thermal energy of SQM
is, in fact, the source of energy for the process of pair creation.

The flux of $e^+e^-$ pairs from the unit surface of SQM is 
\cite{U01a}

\begin{equation}
f_\pm\simeq 10^{39}\left({T_{_{\rm S}}\over 10^9~{\rm K}}
\right)^3\exp \left[{-11.9\left({T_{_{\rm S}}\over 10^9~{\rm K}}
\right)^{-1}}\right]J(\zeta)\,\,\,{\rm cm}^{-2}\,{\rm s}^{-1}\,,
\end{equation}

\noindent
where

\begin{equation}
J(\zeta )={1\over 3}{\zeta^3\ln \,(1+2\zeta ^{-1})\over 
(1+0.074\zeta )^3}+
{\pi^5\over6}{\zeta^4\over (13.9 +\zeta)^4}\,,\,\,\,\,\,\,\,\,\,\,
\zeta = 2\sqrt{{\alpha\over \pi}}{\varepsilon_{_{\rm F}}\over k
T_{_{\rm S}}}\simeq 0.1{\varepsilon_{_{\rm F}}\over k
T_{_{\rm S}}}\,,
\end{equation}

\noindent
and $\alpha = e^2/\hbar c=1/137$ is the fine structure constant.

The energy flux from the unit surface of SQM in $e^+e^-$
pairs created by the Coulomb barrier is $F_\pm\simeq
\varepsilon_\pm f_\pm$, where $\varepsilon_\pm\simeq m_{\rm e}c^2+
k_{_{\rm B}}T_{_{\rm S}}$ is the mean energy of created particles 
\cite{U98}. Figure~1 shows the ratio of the SQM surface emissivity   
in $e^+e^-$ pairs to the blackbody surface emissivity, 
$\xi_\pm =F_\pm /F_{_{\rm BB}}$, versus the surface temperature
$T_{_{\rm S}}$. Creation of $e^+e^-$ pairs by the Coulomb
barrier is the main mechanism of thermal emission from the surface
of SQM at $8\times 10^8 < T_{_{\rm S}} < 5\times 
10^{10}$~K, while the equilibrium radiation dominates at
extremely high temperatures, $T_{_{\rm S}}>5\times 10^{10}$~K.

\subsection{The thermal luminosity of a hot bare strange star} 

At $T_{_{\rm S}} > 8\times 10^8$~K, when the thermal emission
from the SQM surface in both equilibrium photons and $e^+e^-$  pairs 
prevail, the total thermal luminosity of a bare strange star is

\begin{equation}
L=L_{\rm eq}+L_\pm = 4\pi R^2 (F_{\rm eq} +F_\pm )\,,
\end{equation} 

\noindent
where $R\simeq 10^6$~cm is the radius of the strange star. Figure~2 
shows the value of $L$ as a function of the surface temperature 
$T_{_{\rm S}}$.

\begin{figure}[htb]
\begin{center}
\epsfig{file=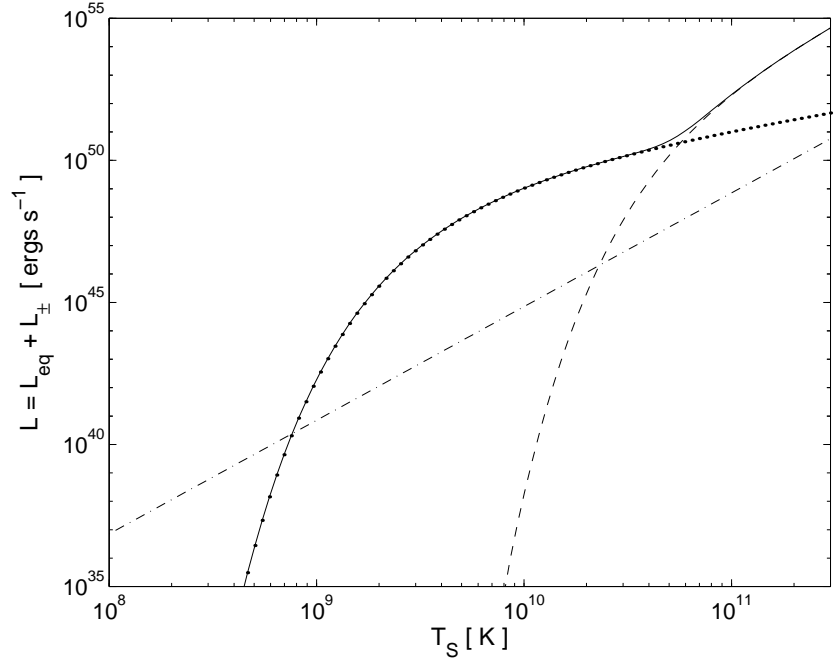,height=3.5in}
\caption{The total luminosity of a bare strange star $L=L_{\rm eq}+
L_\pm$ (solid line), where $L_{\rm eq}$ and $L_\pm$ are the
luminosities in thermal equilibrium photons (dashed line)
and $e^+e^-$ pairs (dotted line), respectively. The upper limit 
on the luminosity in non-equilibrium photons, 
$L_{\rm neq} < 10^{-4}4\pi R^2F_{_{\rm BB}}$, is shown 
by the dot-dashed line.}
\label{fig:luminosity}
\end{center}
\end{figure}

At $T_{_{\rm S}} > 8\times 10^8$~K the luminosity 
in $e^+e^-$ pairs created by the Coulomb barrier at the SQM surface 
is very high,  $L_\pm > 10^{40}$ ergs~s$^{-1}$ (see Fig.~2), 
that is at least four orders of magnitude higher than 

\begin{equation}
L_\pm ^{\rm max}\simeq 4\pi m_{\rm e}c^3R/\sigma_{_{\rm T}}\simeq
10^{36}\,\,{\rm ergs~s}^{-1}\,,
\end{equation}

\noindent
where $\sigma_{_{\rm T}}$ is the Thomson cross-section.
In this case, the time-scale $t_{\rm ann}\sim (n_\pm
\sigma_{_{\rm T}}c)^{-1}$ for annihilation of $e^+e^-$
pairs is much shorter than the time-scale $t_{\rm esc}\sim
R/c$ for their escape, $t_{\rm ann}/t_{\rm esc}\simeq 
L^{\rm max}_\pm/L_\pm <10^{-4}\ll 1$,
and $e^+e^-$ pairs outflowing from
the stellar surface mostly annihilate in the vicinity 
of the strange star, $r\sim R$ (e.g., \cite{B99}).  
The luminosity in $e^+e^-$ pairs at the distance $r\gg R$
cannot be significantly more than $L_\pm^{\rm max}$.
Therefore, far from a bare strange star with the surface temperature
$T_{_{\rm S}} > 8\times 10^8$~K the photon luminosity 
dominates irrespective of $T_{_{\rm S}}$ and practically 
coincides with the total luminosity given by equation (6).
At $T_{_{\rm S}} < 8\times 10^8$~K the total luminosity  
$L=L_\pm + L_{\rm neq}$ is somewhere between
$\sim 4\pi 10^{-4}R^2F_{_{\rm BB}}$ and $\sim L_\pm$.

Till now, we assumed tacitly that SQM at the surface of the strange star
is in the normal (nonsuperconducting) state.
Recently, it was argured that SQM may be a color superconductor if its
temperature is below some critical value (for a review, see 
\cite{ARW98}). In the classic BCS model, the critical temperature is
$T_c \simeq 0.57 \Delta_0 /k_{_{\rm B}}$, where $\Delta_0$ 
is the energy gap at zero temperature. The value of $\Delta _0$ 
is in the range from $\sim 0.1-1$~MeV \cite{BL84}
to $\sim 50-10^2$~MeV \cite{ARW98}. Color superconductivity can 
suppress the nonequilibrium radiation discussed in \cite{CHS91} 
significantly (if not completely). In this case,
equation (6) may be used at $T_{_{\rm S}}<8\times 10^8$~K as well. 
If SQM at the stellar surface is 
a color superconductor in the color-flavor locked (CFL) 
phase the process of $e^+e^-$ pair creation at the SQM surface
may be turned off at $T_{_{\rm }}\ll T_c$. This is because cold
SQM in the CFL phase is electrically neutral, and no electrons
are required and none are admitted inside CFL
quark matter \cite{RW01}. 

The energy spectrum of photons far from the strange star depends on the
total thermal luminosity. At $L > 10^{43}$~ergs~s$^{-1}$, 
the photon spectrum is nearly blackbody with the temperature
$T\simeq T_0(L/ 10^{43}~{\rm ergs~s}^{-1})^{1/4}$, where
$T_0\simeq 2\times 10^8$~K \cite{P90}. For intermediate luminosities,
$10^{42}<L <10^{43}$~ergs~s$^{-1}$, the effective
temperature of photons is more or less constant, $T\sim T_0$ \cite{AMU}.
At $L_{\rm th}< 10^{42}$ ergs~s$^{-1}$, the hardness 
of the photon spectrum increases when $L$ 
decreases. This is because photons that form in annihilation of
$e^+e^-$ pairs cannot reach thermodynamical equilibrium 
before they escape from the strange star vicinity. 
When the photon luminosity decreases from 
$\sim 10^{42}$ ergs~s$^{-1}$ to $\sim 10^{36}$ ergs~s$^{-1}$,
the mean energy of photons increases from $\sim 100$~keV 
to $\sim 500$~keV while the spectrum 
of photons changes eventually into a very wide $(\Delta E
/E\sim 0.3)$ annihilation line of energy $E\sim 500$~keV \cite{AMU}.
Such a behavior of photon spectra offers
a good observational signature of hot bare strange stars.
Super-Eddington luminosities are another finger print of such stars.

\section{Thermal emission from non-bare strange stars}

"Normal" matter (ions and electrons) may be at the quark surface of
strange stars. The ions in the inner layer are supported against 
the gravitational attraction to the underlying strange star 
by a very strong electric field of the Coulomb barrier.
There is an upper limit to the amount of 
normal matter at the quark surface, $\Delta M\leq 10^{-5}M_\odot$ 
\cite{AFO86,GW92}. Such 
a massive envelope of normal matter with $\Delta M\sim 10^{-5}M_\odot$
completely obscures the quark surface. However, a strange star 
at the moment of its formation is very hot. The temperature in 
the interior of a nascent strange star is expected to be as high as a 
few $\times 10^{11}$ K \cite{HPA91}. The rate of mass ejection from an 
envelope of such a hot strange star is very high \cite{WB92}. Besides, 
the high surface temperature leads to a considerable reduction of the 
Coulomb barrier, which favors the tunneling of nuclei toward the quark 
surface \cite{KWWG95}. Therefore, it is natural to expect that 
in a few seconds after formation of a strange star
the normal-matter envelope is either blown away by radiation pressure
or quarkonized, and the stellar surface is completely bare.
The SQM surface remains bare until the thermal luminosity of
the strange star is more than the Eddington limit, $L
> L_{\rm Edd}\simeq 1.3\times 10^{38}(M/M_\odot )$ ergs~s$^{-1}$.

\subsection{Low-mass normal-matter atmospheres}

At $L<L_{\rm Edd}$ the normal-matter atmosphere forms because 
of gas accretion onto the strange star. The presence of the  
atmosphere may restore the ability of the stellar surface to radiate
soft photons (this is like painting with black paint on a silver 
surface). 

The strange star acts on the atmosphere as a heat reservoir. 
At $T_{_{\rm S}}>10^7$~K when the hot gas emits mainly due to 
bremsstrahlung radiation, the thermal structure of the 
low-mass normal-matter atmosphere 
and its photon radiation were considered in \cite{U97} 
by solving the heat transfer problem with $T=T_{_{\rm S}}$ as 
a boundary condition at the inner layer. 
It was shown that if the atmosphere mass $\Delta M$ is smaller than

\begin{equation}
\Delta M_1\simeq 7\times 10^{11}{A\over Z^{2}}
\left({T_{_{\rm S}}\over 10^8\,
{\rm K}}\right)^{3/2}\left({R\over 10^6\,{\rm cm}}\right)^2\,\,\,
{\rm g}\,,
\label{DM}
\end{equation}

\noindent
the atmosphere is nearly isothermal, and its photon luminosity is 

\begin{equation}
L_{\rm a}\simeq  {4\times 10^{33}Z^3
\over A(1+Z)}
\left({R\over 10^6\,{\rm cm}}\right)^{-4}
\left({M\over M_\odot}\right)
\left({T_{_{\rm S}}\over 10^8\,{\rm K}}\right)^{-1/2}
\left({\Delta M\over 10^{12}\,{\rm g}}\right)^2\,\,\,\,{\rm ergs~s}^
{-1}\,,
\label{L2}
\end{equation}

\noindent where $A$ is the mass number of ions and $Z$ is their
electrical charge.

At $\Delta M_1 < \Delta M < \Delta M_2$, convection develops in the 
atmosphere, and the photon luminosity is $\tilde L_{\rm a}=
4\gamma L_{\rm a}/(3\gamma +1)$, where 

\begin{equation}
\Delta M_2\simeq {4\times 10^{12}A\over Z^2\mu^{1/2}}
\left({T_{_{\rm S}}\over 10^8\,
{\rm K}}\right)\left({R\over 10^6\,{\rm cm}}\right)^2\,\,\,\,
{\rm g}\,,
\label{DM2}
\end{equation}

\noindent 
$\mu$=$A/(1+Z)$ is the mean molecular weight,
and $\gamma$ is the ratio of the 
specific heats at constant pressure and at constant volume
\cite{U97}. For a rarefied  totally-ionized plasma we have
$\gamma = {5/3}$ and $\tilde L_{\rm a}=
(10/9)L_{\rm a}$. The difference between $L$ and $\tilde L$
is within the accuracy of our calculations which is $\sim 20$\%.

At $\Delta M>\Delta M_2$, both thermal conductivity and
convection are not able to account for the cooling of
atmospheric matter, and
a thermal instability develops in the atmosphere \cite{U97}. As a
result, the atmosphere cannot be in hydrostatic
equilibrium during a time larger than the characteristic 
cooling time, and it has to be strongly variable on a timescale of
$\sim (10^{-4}-10^{-3})(T_{_{\rm S}}/10^8\,{\rm K})^{1/2}$ s.
This variability of the strange-star atmosphere and its photon
luminosity are known poor. Most probably, at $\Delta M>\Delta M_2$
the tendency of the photon luminosity to increase with
increase of $\Delta M$ holds up to $L_{\rm a} \simeq L_{\rm Edd}$
if $T_{_{\rm S}}> 3\times 10^7$~K \cite{U97}.

The photon emission from the low-mass normal-matter atmosphere of
a hot ($T_{_{\rm S}}> 3\times 10^7$~K) strange star is hard.
The spectrum of this emission is similar to the spectrum of thermal
emission of optically thin plasma at $k_{_{\rm B}}T$ up to
$\sim 10^2$~keV \cite{U97}. This differs 
significantly from the photon emission of neutron stars. 

\subsection{Massive normal-matter envelopes with 
$\Delta M\sim 10^{-5}M_\odot$}

 If the age of a neutron star is $t > 10^2$ yr,
the stellar interior may be divided into two regions: the isothermal 
core with density $\rho >\rho_e\sim 10^{11}$ g cm$^{-3}$ and
the outer envelope with $\rho<\rho_e$ (e.g., \cite{GPE1983}).
Since the density of the normal-matter envelope with  
$\Delta M\sim 10^{-5}M_\odot$ at the quark surface of a strange star
is about $\rho_e$, the temperature variation between the
quark surface and the surface of the normal-matter envelope
is more or less the same as the core-to-surface 
temperature variation of a neutron star for a fixed
temperature at the stellar center. The cooling behavior 
of the quark core of strange stars depends on many factors and
may be more or less similar to the
cooling behavior of the isothermal core of neutron stars 
\cite{SWWG96}. Therefore, from observations of thermal
X-ray emission from not too young ($t>10^2$~yr) compact objects it is
difficult to distinguish strange stars with massive ($\Delta M\sim 
10^{-5}M_\odot$) normal-matter envelopes from neutron stars 
(cf. \cite{P91}).

\section{Soft $\gamma$-ray repeaters may be bare strange stars}

Bare strange stars can radiate at the luminosities greatly exceeding 
the Eddington limit (see \S 2). The mean energy of radiated photons 
is a few ten keV or higher. 
Therefore, bare strange stars are reasonable candidates
for soft $\gamma$-ray repeaters (SGRs) that are the sources of 
brief ($\sim 0.1$~s), intense $[\sim (10^3-10^4)L_{\rm Edd}]$ 
outbursts with soft $\gamma$-ray spectra (for a review on SGRs, see 
\cite{K95}). 
The bursting activity of SGRs may be explained by fast heating 
of the SQM surface of bare strange stars up to the temperature of 
$\sim (1-2)\times 10^9$~K (see Fig.~2) and its subsequent 
thermal emission \cite{U01a,U01b}. The heating mechanism may be either 
impacts of comets onto bare strange stars  
\cite{U01b,ZXQ00} or fast decay of superstrong ($\sim 10^{14}-
10^{15}$~G) magnetic fields \cite{U84}.

Two giant flares were observed on 5~March 1979 and 27 August 1998
from SGR 0526-66 and SGR 1900+14, respectively. The peak luminosity
of these remarkable flares was as high as $\sim 10^{45}$ ergs~s$^{-1}$,
7 orders of magnitude in excess of the Eddington limit for a solar-mass 
object \cite{FKL96}. This luminosity is about ten times higher than 
the luminosity of our Galaxy. Recently, it was shown that the light 
curves of the two giant outbursts may be easily explained in the 
following model \cite{U01b}. A comet-like object with the mass
$M_c\sim 10^{25}$~g falls onto a bare strange star. 
The total duration of the accretion is $\Delta t \sim 10^2-10^3$~s. 
The accreted matter sinks into the strange star and quarkonizes
\cite{AFO86}. During the accretion, $t<\Delta t$,
the surface layers of the strange star are heated, while their 
thermal radiation is completely suppressed by the falling matter. 
The total thermal energy accumulated in the surface layers at the
moment $t=\Delta t$ is $Q\simeq 0.1  M_cc^2\sim 10^{45}$~ergs.
When the accretion is finished and the strange star vicinity is
transparent for radiation, some part of the energy $Q$ is emitted 
from the quark surface and observed as a giant burst. Figure~3 shows 
the light curve expected in this model for $Q=9.2\times 10^{44}$~ergs 
and $\Delta t=370$~s \cite{U01b}. This light curve is in good 
agreement with the light curve observed for the 5 March 1979 event
\cite{FKL96}. The spectrum of this event may be also explained 
by the thermal emission from the strange star \cite{AMU}.

\begin{figure}[htb]
\begin{center}
\epsfig{file=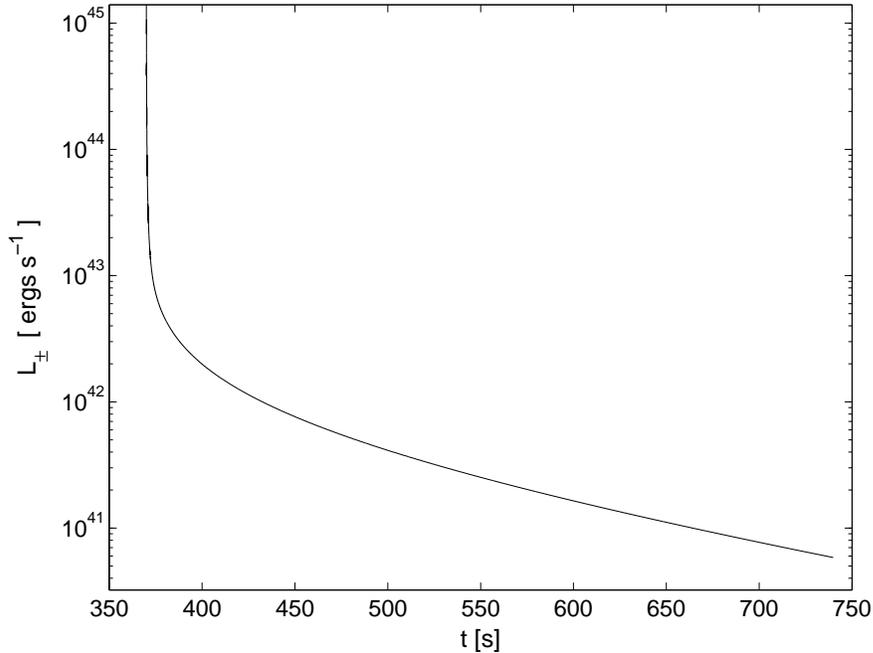,height=3.5in}
\caption{The light curve expected for 
$Q=9.2\times 10^{44}$~ergs and $\Delta t=370$~s.
}
\label{fig:lightcurve}
\end{center}
\end{figure}

\noindent
The light curve of the 27 August 1998 event may be fitted 
fairly well in our model for $Q=6.4 \times 10^{44}$~ergs and 
$\Delta t =270$~s \cite{U01b}.

One of the sources of matter that falls onto 
a strange star producing a SGR could be debris formed in collisions 
of planets orbiting the star in nearly coplanar orbits \cite{KTU94}. 
In this particular model, there appear two typical masses ($\sim 
10^{25}$~g and $\sim 10^{22}$~g) available 
for prompt infall. Accretion of comet-like objects with the masses of 
$\sim 10^{25}$~g  and $\sim 10^{22}$~g may result in 
the giant and typical flares of SGRs, respectively.

It is worth to note that
the distribution of temperature in the surface
layers at the moment $t=\Delta t$, when the accretion is just 
finished and the powerful radiation from the stellar surface 
just starts, completely determines the 
subsequent radiation from the strange star at $t\geq \Delta t$.
If the surface layers of a bare strange star are heated very fast 
($< 10^{-3}$~s) to this temperature distribution  
by any other mechanism, for example by decay of superstrong 
($\sim 10^{14}-10^{15}$~G) magnetic fields \cite{U84}, the 
light curve of the subsequent radiation coincides with the light
curve shown by Figures~3.

Recently, the response of a bare strange star to the energy input onto
the stellar surface was studied numerically \cite{U01c}. In these
simulations, the energy input started at the moment $t=0$, and
it was spherical and steady at $t\geq 0$. A wide range 
of the rate of the energy input was considered,
$10^{38}$~ergs~s$^{-1}\leq L_{\rm input}\leq 10^{45}$~ergs~s$^{-1}$. 
The rise time of the thermal radiation from the strange star 
was calculated for both normal and superconducting SQM.
For giant outbursts ($L_{\rm input}\sim 10^{45}$ 
ergs~s$^{-1}$), the rise time is $ < 10^{-3}$~s 
irrespective of whether SQM is a superconductor or
not. This time is consistent with available data on the two
giant outbursts. For typical outbursts ($L_{\rm input}\sim 10^{41}-
10^{42}$~ergs~s$^{-1}$) the rise time is $\sim 10^2-10^4$~s for
SQM in the normal state and $\sim 10^{-1}-10^{-3}$~s  for SQM
in the superconducting state with the energy gap $\Delta_0 \geq 1$ MeV.
Therefore, for typical outbursts the observed rise times ($\sim
10^{-1}-10^{-3}$~s) may be explained in our model only if SQM is a
superconductor with the energy gap of more than
$\sim 1$~MeV.

\bigskip
This work was supported by the Israel Science Foundation of
the Israel Academy of Sciences and Humanities.

\def\Discussion{
\setlength{\parskip}{0.3cm}\setlength{\parindent}{0.0cm}
     \bigskip\bigskip      {\Large {\bf Discussion}} \bigskip}
\def\speaker#1{{\bf #1:}\ }
\def\endDiscussion{}

\Discussion

\speaker{A. Thampan (IUCAA)}  Won't general relativity modify
(qualitatively) the temperature profile ($T$ versus $x$) that 
you have got?

\speaker{Usov} The effects of general relativity do not change
qualitatively the distribution of temperature in the surface 
layers of the strange star. These effects can lead only to 
rather small ($\sim 20$\%) corrections. In our calculations,
the effects of general relativity were ignored
because many input parameters (for example,
the thermal emission from the SQM surface) are known
within a factor of 2 or so.

\speaker{D.K. Hong (Pusan National University)} In the case of SQM,
why the rise time does not change much as the energy gap changes a lot?

\speaker{Usov} This is because when the energy gap is higher than
about 1~MeV both the specific heat of the quark subsystem of SQM
and its thermal conductivity are strongly suppressed. In this case, 
the heat transport is mostly determined by the electron subsystem, 
and it practically does not depend on the energy gap.

\speaker{J.E. Horvath (Sao Paulo University)} Did you attempt
spectral comparisons of the model with the outburst of
SGR 1900+14 Aug. 9? In that case the light curve has shown
evidence for several periods $\sim $ fractions of a second.
Is there any "natural" room for them in this model?

\speaker{Usov} I have compared the theoretical and observed spectra 
for the 5 March 1979 outburst and found that they are consistent with 
each other. Since the spectra of other outbursts do not differ
qualitatively, I think that these spectra may be explained  
as well. Our consideration of the short-time structure of the light 
curves is just started, and we have no even preliminary results yet.

\endDiscussion
 
\end{document}